\documentclass[%
 aip,
 amsmath,amssymb,
 reprint,%
]{revtex4-2}
\usepackage{graphicx}% Include figure files
\usepackage{dcolumn}% Align table columns on decimal point
\usepackage{bm}% bold math
\usepackage[utf8]{inputenc}
\usepackage[T1]{fontenc}
\usepackage{mathptmx}
\usepackage{etoolbox}
\makeatletter
\def\@email#1#2{%
 \endgroup
 \patchcmd{\titleblock@produce}
  {\frontmatter@RRAPformat}
  {\frontmatter@RRAPformat{\produce@RRAP{*#1\href{mailto:#2}{#2}}}\frontmatter@RRAPformat}
  {}{}
}%
\makeatother
\begin{document}
                                                                                                                                                                                                                                                                                                                                                                                                                                                                                                                                                                                                                                                                                                                                                                                                                                                                                                                                                                                                                                                                                                                                                                                                                                                                                                                                                                                                                                                                                                                                                                                                                                                                                                                                                                                                                                                                                                                                                     
\title{Dynamic characteristics of terahertz hot-electron graphene FET bolometers: effect of electron cooling in channel and at side contacts
}
\author{V.~Ryzhii$^{1}$,   C.~Tang$^{1,2}$, T. Otsuji$^1$, M.~Ryzhii$^{3}$, V.~Mitin$^4$,  and  M. S. Shur$^5$ 
}
\address{
$^1$Research Institute of Electrical Communication,~Tohoku University,~Sendai~ 980-8577,
Japan\\
$^2$Frontier Research Institute for Interdisciplinary Sciences,
Tohoku University, Sendai 980-8578, Japan\\
$^3$Department of Computer Science and Engineering, University of Aizu, Aizu-Wakamatsu 965-8580, Japan\\
$^4$Department of Electrical Engineering, University at Buffalo, SUNY, Buffalo, New York 14260,~USA\\
$^5$Department of Electrical,Computer, and Systems Engineering,\\ Rensselaer Polytechnic Institute,~Troy,~New York~12180,\\ USA\\
}

\begin{abstract}
We analyze the operation of the hot-electron FET bolometers with the graphene
channels (GCs) and the  gate barrier layers (BLs). Such bolometers use the thermionic emission of the hot electrons heated by incident modulated THz 
radiation. The hot electron transfer 
from
the GC into the metal gate. As the  THz detectors, these bolometers
can operate at room temperature.  We show that the response and ultimate modulation
frequency of the GC-FET bolometers are determined by the efficiency of the hot-electron
energy transfer to the lattice and the GC side contacts
due to the 2DEG lateral thermal conductance. The dependences of these mechanisms on
the band structure and geometrical parameters open the way for the GC-FET bolometers
optimization, in particular, for the enhancement of the maximum modulation frequency.
\end{abstract}

\maketitle\section{Introduction}

The specific  band alignment in  metal/black-P$_x$As$_{1-x}$ layer/graphene  structures~\cite{1,2} enables 
an enhanced thermionic electron and hole thermionic emission between  the graphene and metal layers.  
Since the absorption of THz   leads to an effective
electron (holes) heating, the field-effect transistors (FETs) based on such structures 
with the  metal gate (MG),  b-P$_{x}$As$_{1-x}$barrier layer (BL), and  graphene channel (GC) 
 can be used
as sensitive bolometric detectors~\cite{3,4,5}. 
The responsivity of the GC-FET
bolometers is determined by the rate of the carrier cooling due to the transfer of their energy
to the GC and the side contacts (source and drain) as well as the thermionic emission~\cite{4,5,6,7} (the Peltier cooling). 
On the other hand, the same processes determine the speed of the bolometric detectors in question. 
The roles of the  effects in question depend on the structural parameters and the temperature.
The plasmonic resonances in the gate GC of the devices under consideration 
can substantially affect the absorption of the impinging radiation and, hence, the detector performance. 
The recently proposed
 GC-FET bolometers with the composite BL~\cite{8,9} have reinforced plasmonic resonances.
Such a composite BL is made of the h-BN layer with a short narrow-gap black-P$_x$As$_{1-x}$ region. 
The latter serves as the electron emission window, through which the hot electrons pass from the GC into the MG. Since the quality of the  h-BN/GC interface  supports  very high electron mobility (see, for example,~\cite{10,11}) and, therefore, a low electron collision frequency, the plasmonic oscillations damping in the GC-FETs can be markedly weaker than in the GC-FETs with black-P$_x$As$_{1-x}$ BLs.
Due to a relatively narrow emission window, the role of the Peltier cooling is diminished.

In this paper, we consider the GC-FETs  with the n-type GC and composite
h-BN/b-P BL and 
analyze  
 the dynamics of the two-dimensional electron gas (2DEG) heating.  We calculate  the GC-FET bolometer's modulation characteristics and the ultimate modulation frequency of the detected incoming radiation signals 
 as  functions of the device structure parameters and the temperature.

\section{GC-FET detector structure and main model equations}
Figure~1(a) schematically shows the cross-section of the GC-FET structure under consideration.
The GC-FET structure incorporates the GC separated from the MC by the composite 
 h-BN/b-P/h-BN  gate BL. 
For the GC-FETs with the Al MG, one can set 
for the  differences between the
bottom of the BL conduction band and the Dirac point in the GC $\Delta_C = 225$~meV,
and the difference in the electron affinities  of the Al MG
and b-P in the central section of the gate BL $\Delta_M = 85$~meV~\cite{12,13,14,15,16,17,18}. 
We assume that the electron Fermi energy in the GC is 
chosen to satisfy the conditions:
$\Delta_C -\mu = \Delta_M$. 
 The lengths of the b-P  central section and each the h-BN side sections are $2L_C$ and $L-L_C$, respectively [$2L_C < 2L$, see Fig.~1(a)].
 Here $2L$ and $2L_C$ are the length of the GC and the length of the central GC section  
covered by the b-P  section of the gate BL. 
Figs.~1(b) and 1(c) show the GC-FET band diagrams at the gate voltage $V_G$ for different GC sections:
in the side regions ($L_C <|x| < L$) and the central region ($|x| < L_C$).
The current
 between the GC and MG in the side sections ($L_C <|x| < L$) is blocked because of the high-energy h-BN barrier.
Since the energy barrier, $\Delta_C$, for electrons in the GC
in the section covered by the b-P is smaller than in the sections with the h-BN gate BL, this section plays the role of the electron emission window.
The band diagram shown in Figs.~1(c) corresponds to the band parameter assumed above. For  these  conditions, the thermionic current from the GC into the MC is associated with the electrons heated by the absorbed THz radiation
in the whole GC flowing over the barrier via the central region.

\begin{figure*}[!t]
\centerline{\includegraphics[width=15.0cm]{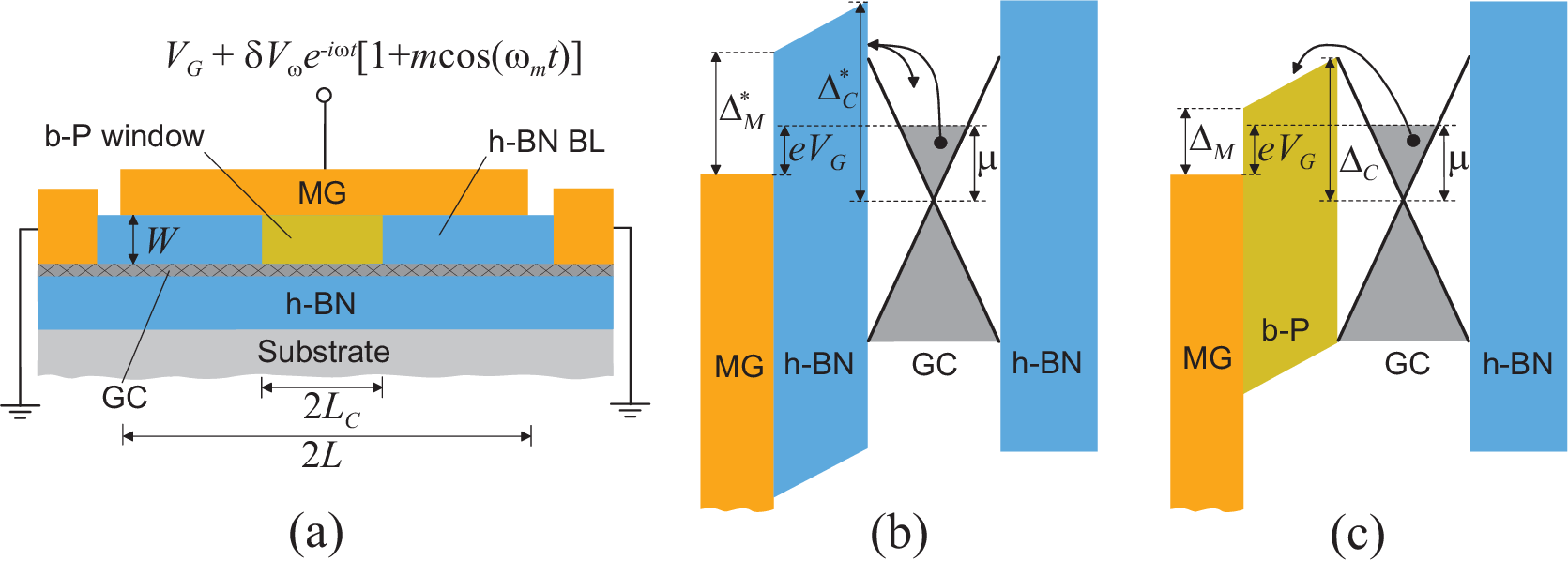}}
\caption{
(a) Cross-section of the GC-FET detector structure with composite BL and  its band diagrams under the applied voltage gate voltage $V_G$ in (b) the side regions ($L_C < |x| <L$) with a high barrier for electrons in GC reflecting
them  and (c) in the central region ($|x| < L_C$) with a moderately high  barrier~(see ~\cite{8,9}).}
\label{fig1}
\end{figure*}

The bias DC gate voltage $V_G$ and the signal ac voltage, $\delta V_{\omega}^{\omega_m} =\delta V_{\omega} \exp(-i\omega t)[1+m\cos(\omega_m t)]$ are applied between the MG and the GC edge contacts.
The signal voltage
is produced by 
the energy flux, $I = I_{\omega}[1 + m\cos(\omega_mt)]$, of
the impinging amplitude-modulated  THz radiation (received by an antenna).
Here $\omega$ is the THz radiation carrier frequency, $m =\delta I_{\omega}/ I_{\omega} < 1$ and $\omega_m$ are the modulation depth
and  frequency ($\omega_m \ll \omega$). 
The absorption of the received amplitude-modulated  THz  radiation 
by the 2DEG  leads to transient heating.
As a result,  the
 electron effective temperature averaged over the period of the carrier signal $2\pi/\omega$ (i.e., over the fast oscillations) is $\langle{\overline T}\rangle = T_0 + \langle \delta T_{\omega}\rangle
 + \langle \delta T_{\omega}^{\omega_m}\rangle $,
 where $T_0$
 is the lattice temperature, $\langle\delta T_{\omega}\rangle$ is the  temperature variation  associated with the heating caused by the carrier signal, and
 $\langle\delta T_{\omega}^{\omega_m}\rangle$ is the slowly varying component associated with the modulation. The symbol $\langle...\rangle$ denotes the averaging over
 the fast oscillations with the characteristic period  $2\pi/\omega_m \gg 2\pi/\omega$). 
The rectified DC and modulation components of the  thermionic current via the b-P section,
$\langle \delta J_{\omega}\rangle$ $\langle \delta J_{\omega}^{\omega_m}\rangle$, can be presented as: 
 
\begin{equation}\label{eq1}
\langle\delta J_{\omega}\rangle = j^{max}H F\int_{-L_C}^{L_C}dx\frac{\langle\delta T_{\omega}\rangle}{T_0},
\end{equation} 

\begin{equation}\label{eq2}
\langle\delta J_{\omega}^{\omega_m}\rangle = j^{max}H F\int_{-L_C}^{L_C}dx\frac{\langle\delta T_{\omega}^{\omega_m}\rangle}{T_0}.
\end{equation} 
Here $j^{max}$ is the maximal value of the current density from the GC, 
(the axis $x$ is directed in the GC plane from one  side contact to the other), $H$ is the GC width (in the in-plane $y$-direction) 
 and

\begin{equation}\label{eq3}
 F = \frac{\Delta_M}{T_0}\exp\biggl(-\frac{\Delta_M}{T_0}\biggr).
\end{equation} 
The exponential factor in the right-hand side of Eq.~(3) has the activation energy for the electrons leaving the GC equal to $(\Delta_C-\mu)$.
The maximal current density
$j^{max}$ 
is estimated as $j^{\max} = e\Sigma/\tau_{\bot}$, where $\Sigma$ and $\tau_{\bot}$ are  the electron density in the GC (both associated with the doping and the gate bias voltage),  the escape time of the electrons with the energy exceeding the barrier height, respectively, and $e$ is the electron charge.. Since the  escape of an electron  from the GC is associated with a significant change in its momentum, we set $\tau_{\bot} = \xi_{\bot}/\nu$, where $\nu$ is the electron scattering frequency (inverse momentum relaxation time) in 2DEG at room temperature and $\xi_{\bot} \sim 1/2\pi$ is a phenomenological parameter (which for
the virtually isotropic 
scattering of the electrons with the energy $\sim \Delta_C$ on acoustic phonons, can be roughly estimated as  $1/2\pi$).

In the GC-FETs under consideration, the thermionic current of the heated electrons passes via the b-P  region. Considering this, Eqs.~(1) and (2) can be transformed to

\begin{equation}\label{eq4}
\langle\delta J_{\omega}\rangle \simeq \frac{\langle\delta T_{\omega}\rangle|_{x=0}}{T_0},
\end{equation} 
\begin{equation}\label{eq5}
\langle\delta J_{\omega}^{\omega_m}\rangle \simeq 2j^{max} L_CH F\frac{\langle\delta T_{\omega}^{\omega_m}\rangle|_{x=0}}{T_0}.
\end{equation}

The slow variations averaged over the fast oscillations  (varying with the characteristic time $2\pi/\omega_m \gg 2\pi/\omega$) of the local  electron temperature, $\langle\delta T_{\omega}^{\omega_m}\rangle$,   are governed by the following 
electron heat transport equation:

\begin{equation}\label{eq6}
- h_e\frac{\partial^2\langle\delta T^{\omega_m}_{\omega}\rangle}{\partial x^2}
+ \frac{\langle\delta T^{\omega_m}_{\omega}\rangle}{\tau_{\varepsilon}}
= \frac{{\rm Re}~\sigma_{\omega}\langle | \delta E_{\omega}|^2\rangle}{\Sigma},
\end{equation} 
\begin{eqnarray}\label{eq7}
c_e\frac{\partial \langle\delta T_{\omega}^{\omega_m}\rangle}{\partial t} 
- h_e\frac{\partial^2\langle\delta T_{\omega}^{\omega_m}\rangle}{\partial x^2}
%\nonumber\\ 
 + 
\frac{\langle\delta T_{\omega}^{\omega_m}\rangle}{\tau_{\varepsilon}}
\nonumber\\
= \frac{{\rm Re}~\sigma_{\omega}\langle | \delta E_{\omega}^{\omega_m}|^2\rangle}{\Sigma}.
\end{eqnarray} 
Here 
 $h_e$ and $c_e $ are thermal conductivity
and the electron thermal capacitance
in the GC, $\tau_{\varepsilon}$ and $\tau_{\bot}$
are the electron energy relaxation time and the try-to-escape time for the electrons emitted via the BL central part, respectively,  
$v_W \simeq 10^8$~cm/s is the characteristic electron velocity in GCs,  $\nu$ is the electron scattering frequency, and
 $\sigma_{\omega} = [\sigma_0 \nu/(\nu-i\omega)]$, where $\sigma_0 = (e^2\mu/\pi\hbar^2\nu)$  is the 2DEG Drude conductivity, and $\delta E_{\omega}^{\omega_m}$ is the ac electric field component in the GC corresponding to the modulated incoming THz radiation. 
 The terms on the left sides of Eqs.~(6) and (7), proportional to $h_e$ and 
 $\tau_{\varepsilon}^{-1}$, describe the electron energy transfer 
to the side contacts due to the electron heat transport along the GC and 
to the phonon system (particularly to  optical phonons). The contribution of the Peltier cooling is disregarded because of the small emission window
($L_C \ll L)$. 
The term on the right-hand side of these equations describes the local power received by the 2DEG in the GC from the incident THz radiation
(per an electron).
Using the general formula for the degenerate 2DEG electron thermal capacitance~\cite{19,20} and the expression for the GC density of states, one can obtain $c_e =(2\pi^2 T_0/3\mu)$. The quantity $h_e = v_W^2/2\nu$, which is in line with the Wiedemann-Franz relation~\cite{21,22}. 
 
Assuming low thermal resistance of the side contacts, the boundary conditions for Eq.~(3) are set to be
\begin{eqnarray}\label{eq8}\langle\delta T_{\omega}\rangle|_{x = \pm L} = 0,\qquad 
\langle\delta T_{\omega}^{\omega_m}\rangle|_{x = \pm L} = 0. 
 \end{eqnarray}

Accounting for the transformation of the THz signal receiving by an antenna
to the AC electric field in the GC under the condition of the plasmonic oscillations excitation, 
 one can obtain the following~\cite{3,5,9} (see, also Refs.~[23,24]):

\begin{eqnarray}\label{eq9}
\langle|\delta E_{\omega}|^2 \rangle = \frac{16}{g\,c}\biggl|\frac{\gamma_{\omega}\sin(\gamma_{\omega}x/L)}{\cos \gamma_{\omega}}\biggr|^2  I_{\omega},
 \end{eqnarray}

\begin{eqnarray}\label{eq10}
\langle|\delta E_{\omega}^{\omega_m}|^2 \rangle = \frac{16}{g\,c}\biggl|\frac{\gamma_{\omega}\sin(\gamma_{\omega}x/L)}{\cos \gamma_{\omega}}\biggr|^2 \delta I_{\omega}\cos(\omega_mt).
 \end{eqnarray}   
 Here $g \simeq 1.64$ is the antenna gain (for a half-wavelength dipole antenna), $c$ is the speed of light in vacuum,
 $\gamma_{\omega} =\pi\sqrt{\omega(\omega+i\nu)}/2\Omega$ and
$\Omega=(\pi\,e/\hbar\,L)\sqrt{\mu\,W/\kappa}$ are the effective wavenumber and the plasmonic frequency, respectively, with
$\kappa$ and $W$ being the dielectric constant of the BL  and its thickness.

Combining Eqs.~(6), (7), (9), and (10), we arrive at

\begin{eqnarray}\label{eq11}
- h_e\frac{\partial^2\langle\delta T_{\omega}\rangle}{\partial x^2}
%\nonumber\\ 
 + 
\frac{\langle\delta T_{\omega}\rangle}{\tau_{\varepsilon}}
\nonumber\\
= 
\beta\frac{v_W^2}{L^2\Omega^2}\biggl(\frac{\hbar\nu}{\mu}\biggr)\frac{\omega}{\sqrt{\omega^2+\nu^2}}\biggl|\frac{\sin(\gamma_{\omega}x/L)}{\cos \gamma_{\omega}}\biggr|^2 I_{\omega},
\end{eqnarray} 

\begin{eqnarray}\label{eq12}
c_e\frac{\partial \langle\delta T_{\omega}^{\omega_m}\rangle}{\partial t} 
- h_e\frac{\partial^2\langle\delta T_{\omega}^{\omega_m}\rangle}{\partial x^2}
%\nonumber\\ 
 + 
\frac{\langle\delta T_{\omega}^{\omega_m}\rangle}{\tau_{\varepsilon}}
\nonumber\\
= 
\beta\frac{v_W^2}{L^2\Omega^2}\biggl(\frac{\hbar\nu}{\mu}\biggr)\frac{\omega}{\sqrt{\omega^2+\nu^2}}\biggl|\frac{\sin(\gamma_{\omega}x/L)}{\cos \gamma_{\omega}}\biggr|^2 
\delta I_{\omega} \cos(\omega_mt),
\end{eqnarray} 
where 
$\beta = \displaystyle\frac{4\pi^2}{137g} \simeq 0.176$ and $\theta$ is the phase shift.
Here we have accounted for  the fine structure constant  $e^2/\hbar\,c = 1/137$.

\section{Output rectified dc and modulation currents}

\begin{figure*}\centering
\includegraphics[width=16.0cm]{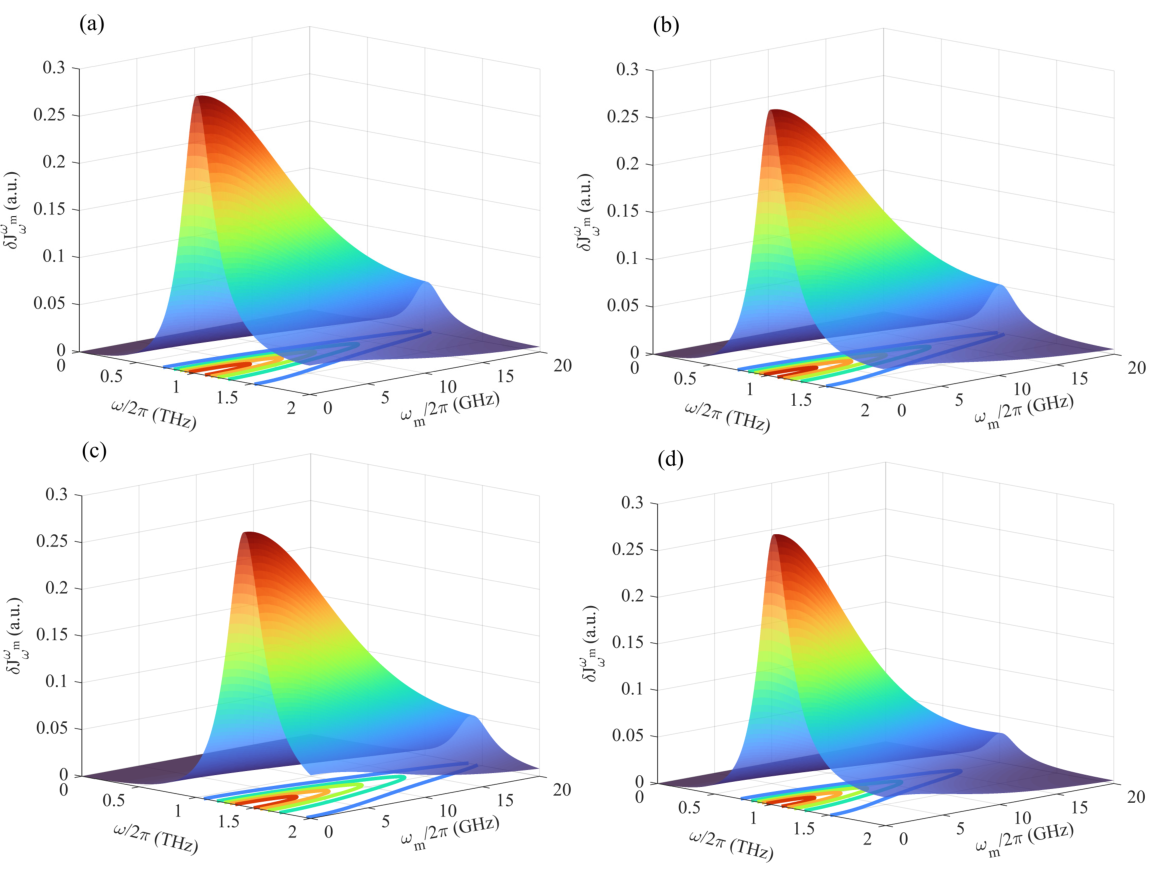}
\caption{
 Amplitude of modulated current $\langle\delta J _{\omega}^{\omega_m}\rangle$  as a function of  signal frequency $\omega/2\pi$  and  modulation frequencies $\omega_m/2\pi$:\\ 
(a)  $2L = 1.4~\mu$m, $\Omega/2\pi = 1.0$~THz, $\nu = 0.5$~ps$^{-1}$, and $\tau_{\varepsilon} = 10$~ps,
%(b)$2L = 1.4~\mu$m, $\Omega/2\pi = 1.0$~THz, $\nu = 1.5$~ps$^{-1}$, and $%\tau_{\varepsilon} = 10$~ps;
%\\
(b) $2L = 1.0~\mu$m, $\Omega/2\pi = 1.4$~THz, $\nu = 1.0$~ps$^{-1}$, and $\tau_{\varepsilon} = 10$~ps,\\
%\\
%(d)$2L = 2.0~\mu$m, $\Omega/2\pi = 0.7$~THz, $\nu = 1.0$~ps$^{-1}$, and $%%%\tau_{\varepsilon} = 10$~ps;
(c) $2L = 1.4~\mu$m, $\Omega/2\pi = 1.0$~THz, $\nu = 1.0$~ps$^{-1}$, %former e
% and $\tau_{\varepsilon} = 7.5$~ps,\\
(d) $2L = 1.4~\mu$m, $\Omega/2\pi = 1.0$~THz, $\nu = 1.0$~ps$^{-1}$, and $\tau_{\varepsilon} = 12.5$~ps.
} 
\label{F2}
\end{figure*}%%%%

In the most interesting frequency range $\omega,  \Omega \gg \nu$, the right-hand sides of Eqs.~(11)
and (12) can be somewhat simplified. This allows to obtain relatively simple and transparent expressions for  $\langle\delta T_{\omega}^{\omega_m}\rangle$ and $\langle\delta T_{\omega}^{\omega_m}\rangle$ in closed analytic  form. 
Such an  approach was verified by the comparison of the results of the analytical and computer modeling.
As shown in Sec.~IV,   the results of the analytical and  computer calculations are very close when
 $\omega,  \Omega \gg \nu$. 

Considering this and  solving simplified versions of 
 Eqs.~(11) and (12) with the boundary conditions given by Eq.~(8), at $\omega,  \Omega \gg \nu$,
 we obtain the following formulas used for the derivation of GC-FET characteristics:

\begin{eqnarray}\label{eq13}
\langle\delta T_{\omega}\rangle|_{x=0} \propto \frac{1}{\mu}\biggl(\frac{v_W\tau_{\varepsilon}}{\nu\,L^2}\biggr) \Pi_{\omega}r_{\omega}I_{\omega},
\end{eqnarray} 
\begin{eqnarray}\label{eq14}
\langle\delta T_{\omega}^{\omega_m}\rangle|_{x=0} \propto \frac{1}{\mu}\biggl(\frac{v_W\tau_{\varepsilon}}{\nu\,L^2}\biggr)
\frac{| \Pi_{\omega}^{\omega_m}|r_{\omega}\cos(\omega_mt +\theta)}{\sqrt{1 +(\omega_m/{\overline \omega}_m)^2}}\delta I_{\omega}.
\end{eqnarray} 
Here
\begin{eqnarray}\label{eq15}
 \Pi_{\omega} \simeq 1 - \frac{1}{1+(\pi\omega/a\Omega)^2}\nonumber\\
 -\biggl[1 - \frac{\cos(\pi\omega/\Omega)}{1+(\pi\omega/a\Omega)^2}\biggr]\frac{1}{\cosh(a)},
\end{eqnarray} 
\begin{eqnarray}\label{eq16}
 \Pi_{\omega}^{\omega_m}\simeq 1 - \frac{1}{1+(\pi\omega/a_m\Omega)^2}\nonumber\\
  -\biggl[1 - \frac{\cos(\pi\omega/\Omega)}{1+(\pi\omega/a_m\Omega)^2}\biggr]\frac{1}{\cosh(a_m)},
\end{eqnarray} 
where  the parameters $a_m =  a\sqrt{1-i\omega_m/{\overline \omega}_m}$,
$ a = L/{\mathcal L} =L\sqrt{2\nu/v_W^2\tau_{\varepsilon}} $ with   ${\mathcal L} = \sqrt{h_e\tau_{\varepsilon}}$, and ${\overline\omega} _m = 1/c_e\tau_{\varepsilon}$ characterizing the 2DEG cooling at the side contacts, 
\begin{eqnarray}\label{eq17}
r_{\omega} = [\sin^2(\pi\omega/2\Omega) + (4\Omega/\pi\nu)^2\cos^2(\pi\omega/2\Omega)]^{-1}
\end{eqnarray} 
is the factor describing the plasmonic resonances, and $\theta$ is a phase shift. The factor $\mu^{-1}$ in the right-hand sides of Eqs.~(13) and (14) appears because the  2DEG conductivity
and density are  $\sigma_{\omega} \propto \sigma_0 \propto \mu$ and  $\Sigma \propto \mu^2$, respectively,  so that
$\sigma_{\omega}/\Sigma \propto \mu^{-1}$. One needs to note that the Fermi energy $\mu$ is assumed  to be fixed to provide a proper band alignment as stated above.

Further, for the rectified dc current $\langle J_{\omega}\rangle$,
 the amplitude of the modulated current $\langle \delta J_{\omega}^{\omega_m}\rangle$, and for the pertinent current responsivities ${\mathcal R}_{\omega}=\langle J_{\omega}\rangle/I_{\omega}$
 and  ${\mathcal R}_{\omega}^{\omega_m}= \langle\delta J_{\omega}^{\omega_m}\rangle/ \delta I_{\omega} $
we obtain using Eqs.~(4), (5), (13), and (14):

\begin{eqnarray}\label{eq18}
 R_{\omega}  \propto  \frac{2L_CF}{\tau_{\bot}}\biggl(\frac{v_W^2\tau_{\varepsilon}}{\nu\,L^2}\biggr)r_{\omega}
| \Pi_{\omega}|,
\end{eqnarray}

\begin{eqnarray}\label{eq19}
 R_{\omega}^{\omega_m} \propto  \frac{2L_CF}{\tau_{\bot}}\biggl(\frac{v_W^2\tau_{\varepsilon}}{\nu\,L^2}\biggr)\frac{r_{\omega}
 | \Pi_{\omega}^{\omega_m}|}{\sqrt{1 +(\omega_m/{\overline \omega}_m)^2}}.
\end{eqnarray}
Accordingly, for the voltage responsivities we obtain
\begin{eqnarray}\label{eq20}
 U_{\omega} = R_{\omega}\rho_L,\qquad
 U_{\omega}^{\omega_m} = R_{\omega}^{\omega_m}\rho_L,
\end{eqnarray}
where $\rho_L \propto 1/L_CF$ is the load resistance (in the GC/MG circuit).

Figure~2 shows the modulated current responsivity ${\mathcal R}_{\omega}^{\omega_m}$ (amplitude of modulated current $\langle\delta J _{\omega}^{\omega_m}\rangle$) versus the signal frequency $\omega/2\pi$  and  modulation frequencies,
$\omega/2\pi$ and 
$\omega_m/2\pi$, calculated using Eq.~(19).

In this and the following figures we assume that
 $W = 10$~nm, $\kappa = 4$,
$\mu = 140$~meV (to fit the condition $\mu = \Delta_C - \Delta_M$,
where for Al MG $ \Delta_C =225$~meV  and $\Delta_M = 85$~meV),
 $\tau_{\varepsilon} = 5-15$~ps,
  $\nu = 0.5 - 1.5$~ps$^{-1}$, 
 and $T_0 = 25$~meV, so that 
 %$\Omega/2\pi = 1.0$~THz and 
 $c_e = 1.175$ (${\overline \omega}_m/2\pi \simeq 9 - 27$~GHz). 
The chosen values of $\tau_{\varepsilon}$ generally correspond to the electron energy relaxation in GCs primarily on optical phonons~\cite{25,26,27,28,29}.
As seen in Fig.~2, the amplitude of the modulated current exhibits   maxima associated with the fundamental plasmonic resonances $\omega \simeq \Omega$. 
 When the signal frequency $\omega$ tends to zero,
the modulation current $\delta J_{\omega}^{\omega_m}$ (as well as $\delta J_{\omega}$) also approaches  zero. 
This is because of the heating electric field along the GC $\delta E_{\omega}$
vanishes in line with Eqs.~(9) and (10).
According to the plots in Fig.~2, the position of  resonant peak shifts with changing GC length $2L$ according to $\Omega \propto L^{-1}$ [compare Figs.~2(a) and 2(b)]. 
Comparing Figs.~2(c) and 2(d), one can see that the variation of the electron energy relaxation time
$\tau_{\varepsilon}$ leads first to higher values of $\delta J_{\omega}^{\omega_m}$ at low modulation frequencies $\omega_m/2\pi$ and to a faster  
decrease in $\delta J_{\omega}^{\omega_m}$ at higher  $\omega_m/2\pi$.
A lowering of  the resonant maxima  with rising  modulation frequency $\omega_m$ is attributed to the weakening of the 2DEG heating when $(\omega_m/{\overline \omega}_m)$ increases

Figures.~3 and 4 show the dependencies of the modulated current peak values   $\langle\delta J _{\Omega}^{\omega_m}\rangle \propto R _{\Omega}^{\omega_m}$ on the electron collision frequency $\nu$
(at fixed electron energy relaxation time $\tau_{\varepsilon}$)
and on $\tau_{\varepsilon}$ (at fixed $\nu$) calculated for different GC length $2L$. To maintain $\Omega/2\pi = 1.0$~THz for different $L$,
 the curves of Figs.~3 and 4 correspond to different values of   the BL $W$ thickness ($\Omega \propto \sqrt{W}/L = const$).

\begin{figure}\centering
\includegraphics[width=9.0cm]{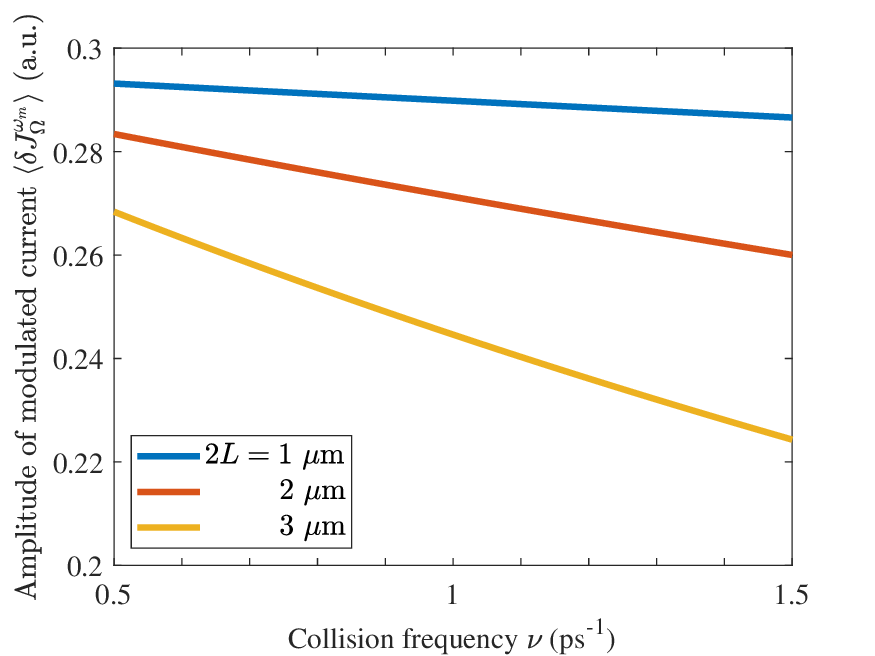}
\caption{
Amplitude of modulated current $\langle\delta J _{\omega}^{\omega_m}\rangle$  as a function of the electron collision frequency $\nu$ for different  GC length $2L$ ($\tau_{\varepsilon} = 10$~ps, $\omega_m/2\pi = 13.5$~GHz).
} 
\label{F3}
\end{figure}

\begin{figure}\centering
\includegraphics[width=9.0cm]{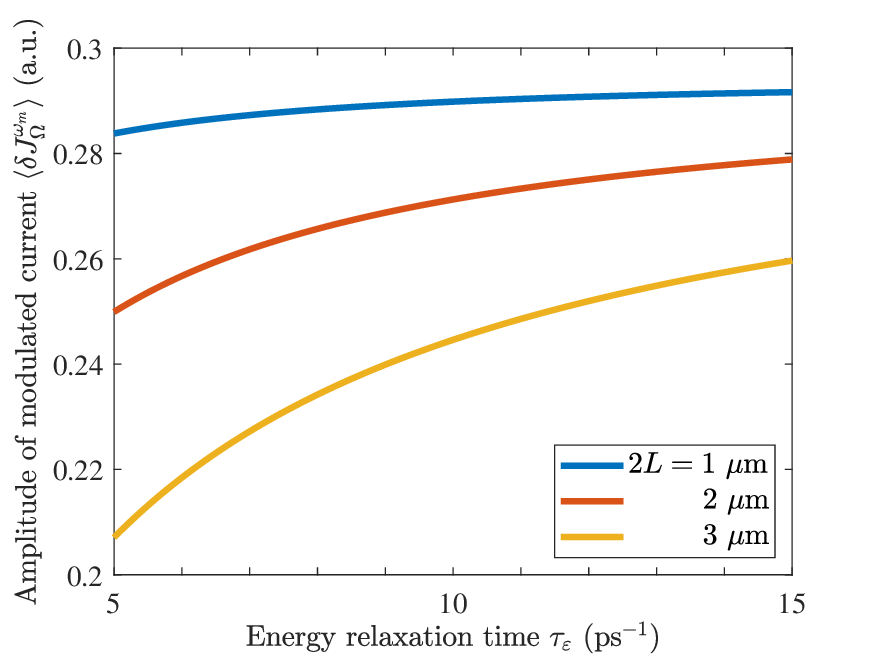}
\caption{
 Amplitude of modulated current  $\langle\delta J _{\omega}^{\omega_m}\rangle
 $ as a function of the electron energy relaxation time  $\tau_{\varepsilon}$ for different  GC length $2L$ ($\nu = 1.0$~ps$^{-1}$, $\omega_m/2\pi = 13.5$~GHz).
} 
\label{F4}
\end{figure}

\section{Modulation characteristic}

Introducing the detector modulation efficiency 
\begin{eqnarray}\label{eq21}
M_{\omega}^{\omega_m} = \frac{  R_{\omega}^{\omega_m}}{  R_{\omega}} =\frac{ U_{\omega}^{\omega_m} }{ U_{\omega}}
%\biggl|\frac{\langle \delta J_{\omega}^{\omega_m}\rangle}
%{\langle\delta J_{\omega}\rangle}\biggr|\biggl/
% \biggl|\frac{\delta I_{\omega}}{I_{\omega}}\biggr|
\end{eqnarray} 
and accounting for Eqs.~(18) and (19), for  the fundamental plasmonic resonance ($\omega = \Omega$) we obtain

 \begin{eqnarray}\label{eq22}                                                                                                        M_{\Omega}^{\omega_m}
 = \frac{1}{\sqrt{1+ (\omega_m/{\overline \omega}_m)^2}}\biggl|\frac{ \Pi_{\Omega}^{\omega_m}}{ \Pi_{\Omega}}\biggr|,
\end{eqnarray}
i.e., accounting for Eqs.~(15) - (17),

\begin{eqnarray}\label{eq23}                                                                                                       
M_{\Omega}^{\omega_m} =
 \frac{1}{\sqrt{1+ (\omega_m/{\overline \omega}_m)^2}}\nonumber\\
\times\Biggl|\frac{1 - \displaystyle\frac{1}{1+(\pi/a_m)^2}
  -\displaystyle\biggl[1 + \frac{1}{1+(\pi/a_m)^2}\biggr]\frac{1}{\cosh(a_m)}}
  { 1 - \displaystyle\frac{1}{1+(\pi/a)^2}
 -\displaystyle\biggl[1 + \frac{1}{1+(\pi/a)^2}\biggr]\frac{1}{\cosh(a)}}\Biggr|.
\end{eqnarray}

Figure~5 shows                                                                                                              the modulation efficiency $M_{\Omega}^{\omega_m}$ versus modulation frequency $\omega_m/2\pi$ for  detectors 
with different GC lengths $2L$ and different values of scattering frequency $\nu$ in the main part of the CC (i.e., covered by the h-BN)  at the plasmonic resonance $\omega = \Omega$. 
 The dashed lines correspond to the first factor in the right-hand sides of Eqs.~(22) and (23) $  {\overline M}_{\Omega}^{\omega_m} = 1/\sqrt{1+(\omega_m/{\overline \omega}_m)^2}$. This factor describes the net efficiency $M_{\Omega}^{\omega_m}$ roll-off with increasing modulation frequency $\omega_m$ associated solely with the 2DES cooling due to the electron energy relaxation 
in the GC. The inset on Figs.~5 shows the dependence  of the second factor in Eqs.~(22) and (23), i.e., $\Theta_m =  \Pi_{\omega}^{\omega_m}/ \Pi_{\omega}$ on $2L$ and $\nu$.
The factor $\Theta_m$ reflects the effect of the 2DEG cooling due to the electron  heat transfer to the side contacts with the absorption of the electron's excessive thermal energy by these contacts (the heat transfer factor).
As seen from  Fig.~5, the modulation efficiency $M_{\Omega}^{\omega_m}$ 
is markedly larger  than ${\overline M}_{\Omega}^{\omega_m}$ (compare the solid and dashed lines). This is attributed to
the fact that   $\Theta_m$ is larger than 
unity as shown in the inset. In the latter case the maximal modulation frequency $\omega_m^{max}$ defined by the relation $M_{\Omega}^{\omega_m^{max}} = 1/\sqrt{2}$
exceeds  ${\overline \omega}_m$. %This is confirmed also by Figs.~6 and 7. 

Figures.~6 and 7 show the maximal modulation frequency $\omega_m^{\max}/2\pi$.
The red curves  in Figs.~6 and 7 correspond to the same set of parameters.

As follows from Figs.~5 - 7, the maximal modulation frequency $\omega_m^{max}/2\pi$ in the GC-FET detectors with the
GC length $2L = (1 - 3)~\mu$m can be about   dozens GHz.

\begin{figure}\centering
\includegraphics[width=9.0cm]{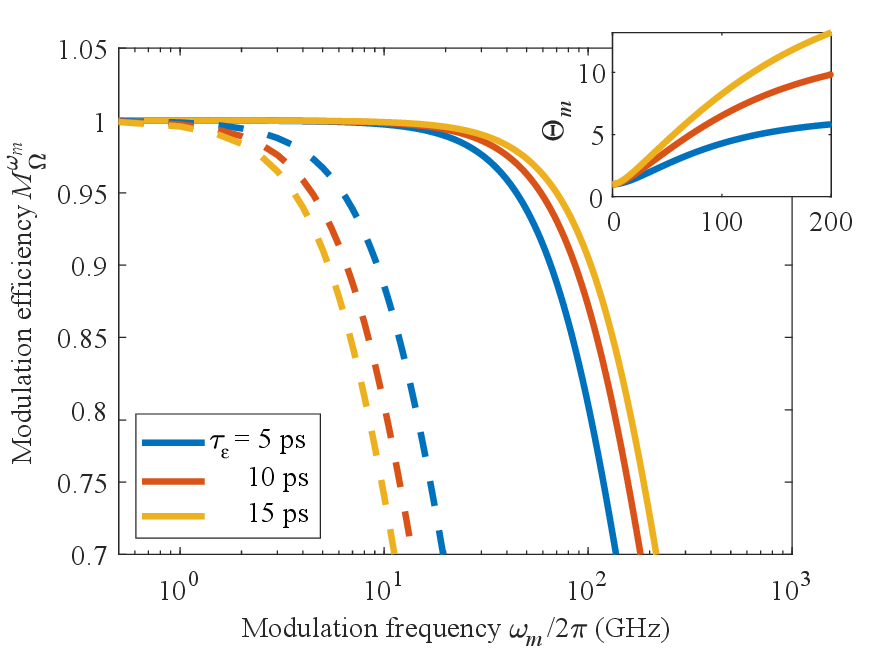}
\caption{Modulation efficiency $M^{\omega_m}_{\Omega}$ at the plasmonic resonance ($\omega = \Omega$) versus the modulation frequency $\omega_m/2\pi$ for 
 different values of electron energy relaxation time $\tau_{\varepsilon}$ 
( $2L = 2.0~\mu$m, $\nu= 1.0$~ps$^{-1})$.
The dashed lines and the inset correspond  to ${\overline M}_{\Omega}^{\omega_m}$ and $\Theta_m $ versus $\omega_m/2\pi$ dependences.}
\label{F5}
\end{figure}%%%%

\begin{figure}[t]\centering
\includegraphics[width=9.0cm]{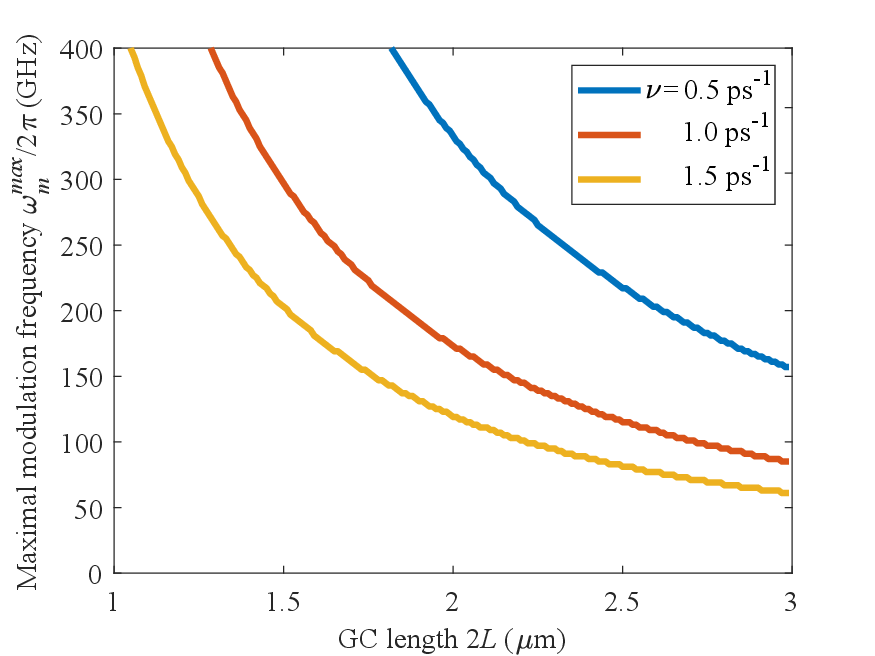}
\caption{Maximal modulation  frequency $\omega_m^{max}/2\pi$
  versus GC length $2L$ for different values of collision frequency $\nu$ and $\tau_{\varepsilon} = 10$~ps.
} 
\label{6}
\end{figure}%%%%

\begin{figure}[t]\centering
\includegraphics[width=9.0cm]{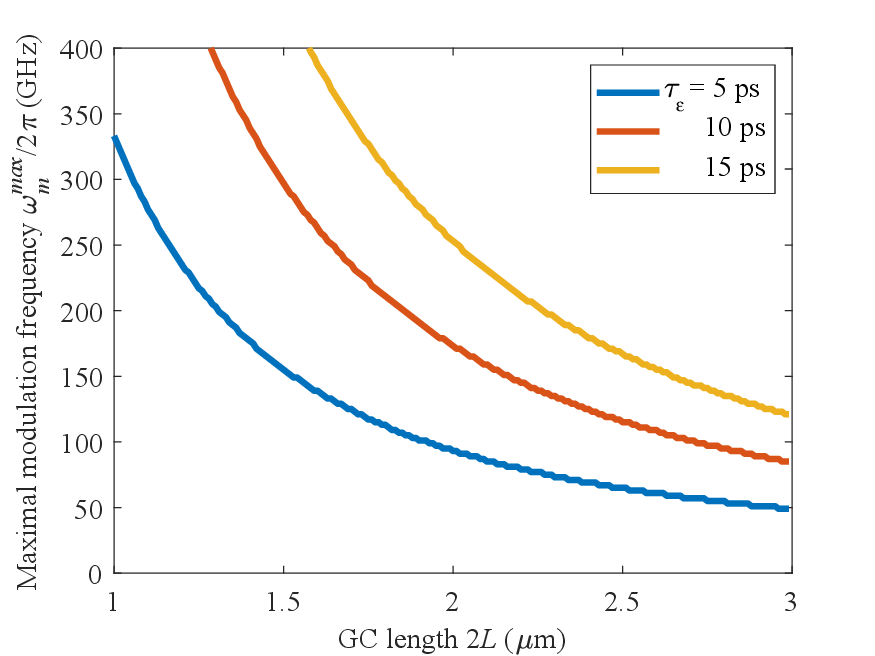}
\caption{The same dependences as in Fig.~6, but for different values of electron energy relaxation time $\tau_{\varepsilon}$ ($\nu = 1.0$~ps$^{-1}$).
} 
\label{7}
\end{figure}%%%%

\section{Comparison of   modulation characteristics
of GC-FETs with composite and uniform BLs}

The characteristics of GC-FETs with the uniform BL  can be obtained from
Eqs.~(1), (2), (11), and (12) setting $L_C = L$. 
In this case, the electron collision frequency $\nu$ and the factors $ \Pi_{\omega}$ and  $ \Pi_{\omega}^{\omega_m}$
should be replaced by $\tilde \nu$,
 $ \tilde{ \Pi}_{\omega}$, and  $ \tilde {\Pi}_{\omega}^{\omega_m}$, respectively. The electron collision frequency in the GC encapsulated in h-BN $\nu$
 is usually smaller than that in the case of the GC sandwiched between the h-BN
substrate and the b-P BL $\tilde \nu$~\cite{30}. 
The consideration of 
the intermediate case $L_C \lesssim L$ leads to rather cumbersome formulas and, 
therefore, will not be studied below. 

Comparing the 
modulated current responsivities,   ${\mathcal R}_{\Omega}^{\omega_m}$
and $\tilde{{\mathcal R}}_{\Omega}^{\omega_m}$, 
and  the modulation efficiencies,  $M_{\Omega}^{\omega_m}$ and 
${\tilde M}_{\Omega}^{\omega_m}$,
of the GC-FET bolometric detectors with that of the composite h-BN/b-P BL given by Eqs.~(19) and (23)  and those of the detectors with the uniform b-P BL 
(which are derived using~\cite{9}) at the plasmonic resonance, we obtain

\begin{eqnarray}\label{eq24}                                                                                                       
{\mathcal R}^{\omega_m}_{\omega} =\frac{ R_{\omega}^{\omega_m}}
{\tilde{ R}_{\omega}^{\omega_m}} = \frac{L_C}{L}\frac{\tilde\nu}{\nu}\nonumber\\
\times
\Biggl|\frac{1 - \displaystyle\frac{1}{1+(\pi/a_m)^2}
  -\displaystyle\biggl[1 + \frac{1}{1+(\pi/a_m)^2}\biggr]\frac{1}{\cosh(a_m)}}{ 1 
 -\displaystyle\biggl[1 + \frac{1}{1+(\pi/\tilde a_m)^2}\biggr]\frac{\tanh(\tilde a_m)}{\tilde a_m}}\Biggr|
\end{eqnarray}
and
\begin{eqnarray}\label{eq27}                                                                                                       
\frac{M_{\Omega}^{\omega_m}}{{\tilde M}_{\Omega}^{\omega_m}} =
\Biggl|\frac{1  
  -\displaystyle\biggl[1 + \frac{1}{1+(\pi/\tilde a)^2}\biggr]\frac{\tanh(\tilde a)}{\tilde a}}
  { 1 
 -\displaystyle\biggl[1 + \frac{1}{1+(\pi/\tilde a_m)^2}\biggr]\frac{\tanh(\tilde a_m)}{\tilde a_m}}\Biggr|\nonumber\\
\times\Biggl|\frac{1 - \displaystyle\frac{1}{1+(\pi/a_m)^2}
  -\displaystyle\biggl[1 + \frac{1}{1+(\pi/a_m)^2}\biggr]\frac{1}{\cosh(a_m)}}
  { 1 - \displaystyle\frac{1}{1+(\pi/a)^2}
 -\displaystyle\biggl[1 + \frac{1}{1+(\pi/a)^2}\biggr]\frac{1}{\cosh(a)}}\Biggr|,
\end{eqnarray}
respectively.
Here
$\tilde a = L\sqrt{2\tilde\nu/v_W^2\tau_{\varepsilon}}$,
$\tilde a_m = \tilde a\sqrt{1-i\omega_m/{\overline\omega}_m}$, 
 hence $\tilde a/a = \tilde a_m/a_m = \sqrt{\tilde\nu/\nu} $ providing that the GCs of  devices of both types have the same length $2L$ and the electron energy relaxation time $\tau_{\varepsilon}$, but different electron scattering frequencies $\nu$ and $\tilde\nu$. 

If the load resistance is chosen to be equal to the GC/MG resistance, i.e.,
inversely proportion to $L_CF$,

\begin{eqnarray}\label{eq28}                                                                                                       
{\mathcal U}^{\omega_m}_{\Omega} = \frac{ U_{\Omega}^{\omega_m}}{{\tilde{ U}}_{\Omega}^{\omega_m}} = \frac{L}{L_C}\frac{ R_{\omega}^{\omega_m}}
{\tilde{ R}_{\omega}^{\omega_m}} \propto \frac{\tilde \nu}{\nu}.
\end{eqnarray}
The latter ratio is independent of $L_C$.

Figs. 8 and 9 show the results of the comparison of the voltage responsivities of the GC-FET detectors with the composite h-BN/b-P/h-BN gate BL and with the uniform b-P gate BL. One can see that the responsivity of the former exceeds that of the latter
when $\nu < {\tilde \nu}$, which corresponds to the reality. This confirms
the advantage of using the composite gate BL in the GC-FET bolometric detectors under consideration.

\begin{figure}[t]\centering
\includegraphics[width=9.0cm]{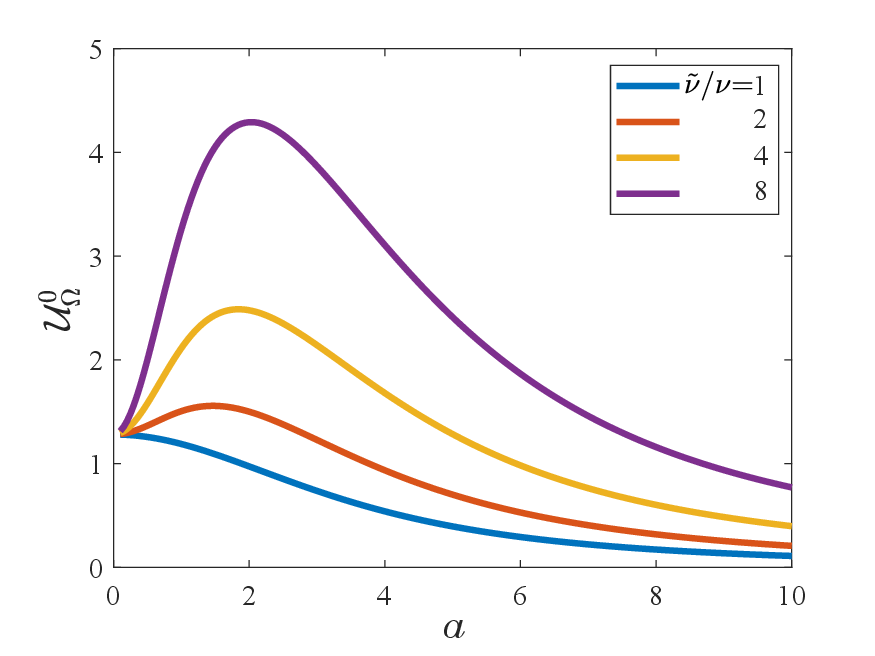}
\caption{
Ratio of  modulation voltage responsivities 
${\mathcal U}^{0}_{\Omega}$ at low modulation frequencies and  plasmonic resonance
($\omega_m \ll \omega_m^{max}$ and $\omega = \Omega$)
as a function of parameter $a = L/{\mathcal L} = L\sqrt{2\nu/v_W^2\tau_{\varepsilon}}$ for different ratios 
and  $\tilde \nu/\nu = (1 - 8)$.
} 
\label{8}
\end{figure}

\begin{figure}[t]\centering
\includegraphics[width=9.0cm]{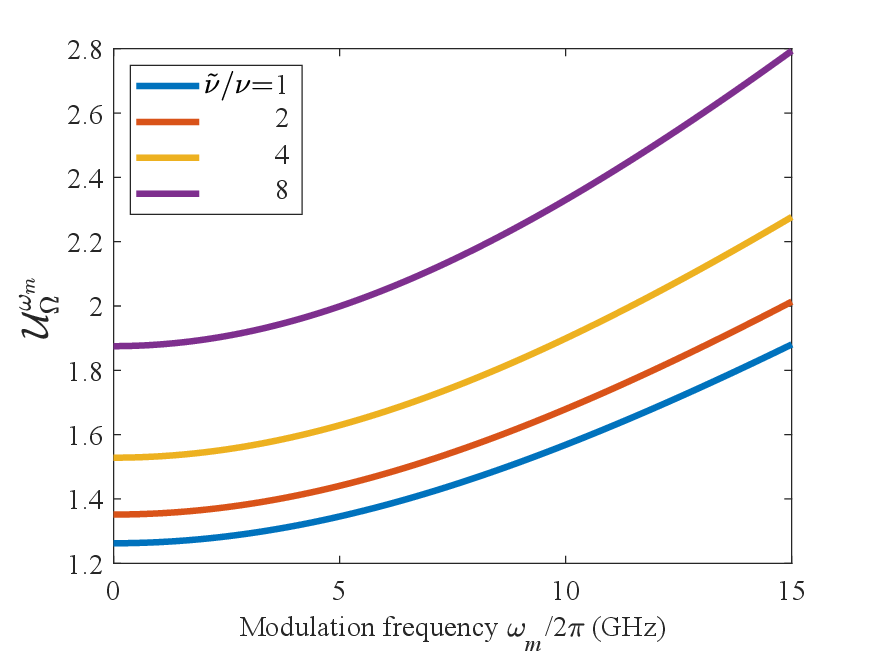}
\caption{
Frequency dependences   
of the voltage responsivities  ratio,  ${\mathcal U}^{\omega_m}_{\Omega}$,
 for  the GC-FET with the composite gate BL (and different $\tilde \nu$) and the GC-FETs with uniform gate: $\nu = 1.0$~ps$^{-1}$
 $L = 1 ~\mu$m and $\tau_{\varepsilon} = 10$~ps.
} 
\label{9}
\end{figure}

\section{Comparison of computer and analytical modeling: verification of  analytical model}

\begin{figure}[b]\centering
\includegraphics[width=9.0cm]{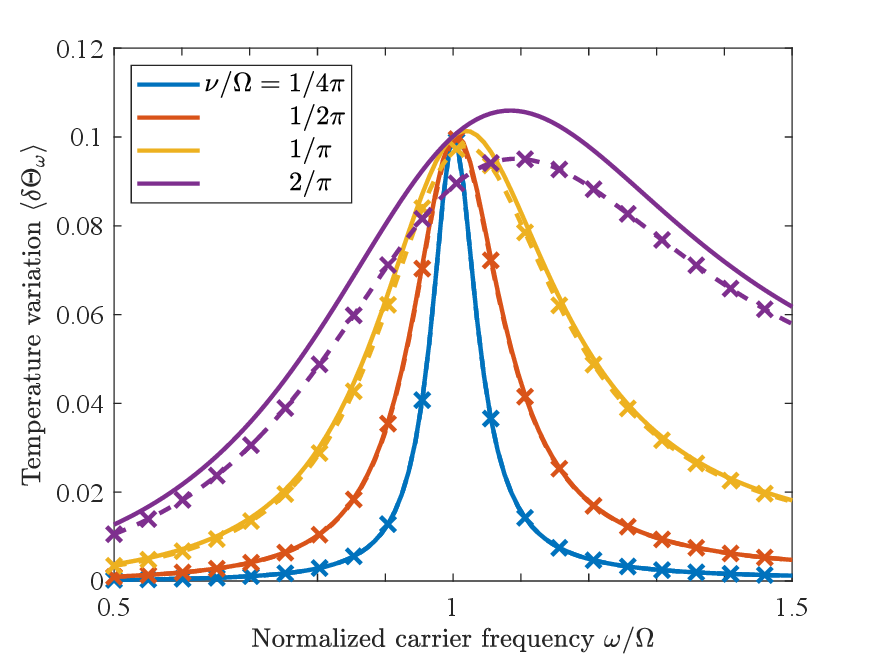}
\caption{
Frequency dependences of the normalized
temperature variations  $\langle\delta \Theta_{\omega}\rangle |_{x=0}$ in the structure center obtained
using computer (dashed lines with markers) and analytical
(solid lines) modeling for different values of $\nu/\Omega$ and $a=1$.
} 
\label{10}
\end{figure}

\begin{figure}[t]\centering
\includegraphics[width=9.0cm]{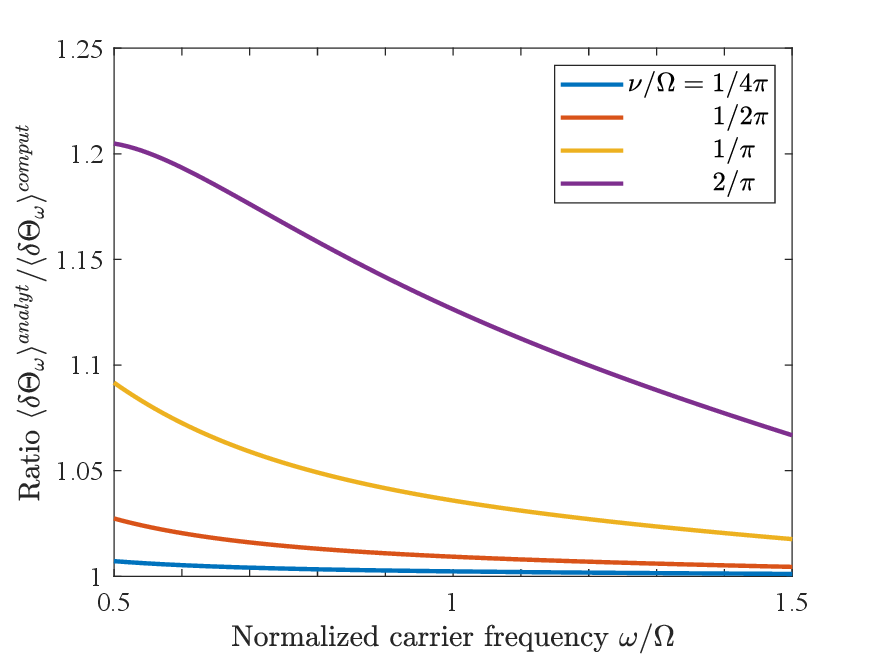}
\caption{Ratios of the normalized temperature variations
$\langle \delta \Theta_{\omega}^{analyt} \rangle|_{x=0}$ 
and $\langle \delta \Theta_{\omega}^{comput} \rangle|_{x=0}$ 
in the structure center as functions of normalized carrier frequency
$\omega/\Omega $ obtained  for different values of $\nu/\Omega$ and $a=1$.
} 
\label{11}
\end{figure}

To verify the accuracy of the analytical results we compare the values of the normalized
electron temperatures, $\langle\delta \Theta_{\omega}\rangle^{comput}$ and $\langle\delta \Theta_{\omega}\rangle^{analyt}$, obtained from the differential Eq.~(11) and using the analytical
formula. These values are defined as
$$
\langle \delta \Theta_{\omega}\rangle^{comput} = \langle \delta T_{\omega}\rangle^{comput} \biggl/\biggl(\frac{32\beta\hbar}{\pi^2\mu}I_{\omega}\biggr)
,$$
$$ \langle \delta \Theta_{\omega}\rangle^{analyt} = \langle \delta T_{\omega}\rangle^{analyt} \biggl/\biggl(\frac{32\beta\hbar}{\pi^2\mu}I_{\omega}\biggr).
$$
As a result, we arrive at the following dimensionless equations, respectively:

\begin{eqnarray}\label{eq27}
- \frac{\partial^2\langle\delta \Theta_{\omega}\rangle^{comput}}{\partial \xi^2}
 + 
a^2\langle\delta \Theta_{\omega}\rangle^{comput}
= 
\biggl(\frac{\pi\nu}{4\Omega}\biggr)^2\frac{\omega}{\sqrt{\omega^2+\nu^2}}\nonumber\\
\times
\biggl|\frac{\sin(\gamma_{\omega}\xi)}{\cos \gamma_{\omega}}\biggr|^2, 
\end{eqnarray} 

\begin{eqnarray}\label{eq28}
\langle\delta \Theta_{\omega}\rangle^{analyt}= \frac{r_{\omega}}{2a^2}\nonumber\\
\times
\biggl\{
1 - \displaystyle\frac{\cos(\pi\,\omega\,x/\Omega\,L)}{1+(\pi \text{\ae}\,\omega/\Omega)^2}
- \displaystyle\biggl[1-\frac{\cos(\pi\,\omega/\Omega)}{1 + (\pi\,\omega/a\Omega)^2}\biggr]
\frac{\cosh(ax/L)}{\cosh(a)}\biggr\}.
\end{eqnarray}
Here $\xi = x/L$, so that the boundary conditions are $\langle\delta \Theta_{\omega}\rangle^{comput}|_{\xi=\pm 1} = \langle\delta \Theta_{\omega}\rangle^{analyt}|_{\xi=\pm 1} = 0$, and the factor $r_{\omega}$ given by Eq~(17).

All equations in this work were numerically calculated with MATLAB (version 9.14.0 R2023a, Natick, Massachusetts: The MathWorks Inc.). Finding the maximal modulation frequency $\omega^{max}_m$ (Figs. 6 and 7), we used the Parallel Computing Toolbox ({\it parfor}) to speed up massive calculations. The differential Eq.~ (27) was solved with a standard MATLAB function {\it dsolve}.

Figure~10 shows that at low ratios of $\nu/\Omega$ ($\nu/\Omega \leq 1/\pi$), the computer and analytical calculations provide
practically distinguishable dependences. However, at a relatively large values of $\nu/\Omega$ (of $\nu/\Omega = 2/\pi$), the distinction is visible reaching about 15$\%$.
This is confirmed also by the plots in Fig.~11. 

Similar conclusion can be made in respect of the results following from Eqs.~(12) and (14)
(with the substitution of parameter $a$ by $a_m$).

Thus, the above comparison of the results of the computer  and analytical models justifies using Eqs.~(13) and (14), which provide the GC-FET characteristics with sufficiently high accuracy.

\section{Comments}

The GC-FET detectors with a larger $2L$ exhibit smaller modulation currents $|\delta J^{\omega_m}_{\Omega}|$(see Figs.~3 and 4). This is because at the same intensity of the impinging radiation,
the amplitude of the signal electric field in the GC $|\delta E_{\omega}|$ decreases with increasing $L$ resulting in a weaker electron heating. As seen from Eqs.~(15)
and (16), $\langle \delta T_{\omega}\rangle \propto \langle \delta T_{\omega}^{\omega_m}\rangle \propto L^{-2}$. However, an increase in $2L$ leads to a diminishing
of the electron energy transfer to the side contact. The trade-off of such factors,
gives rise to a relatively weak dependence of  the modulation current of the GC length. This, in particular, follows from the comparison of $\langle\delta J^{\omega_m}_{\Omega}\rangle$ at $2L = 1~\mu$m and $2L = 3~\mu$m shown in Figs.~3 and 4.

Above, calculating the GC-FET characteristics,  we assumed the room temperature
operation. Lowering  the working temperature might lead to a marked change of the GC-FETs performance as  bolometric detectors.
At lower   the lattice temperatures $T_0$,  $\nu$, and $F$ decrease, while
 $\tau_{\varepsilon }$ simultaneously becomes larger.
In particular, at not too low temperatures, 
the electron energy relaxation time and the electron scattering frequency  are determined by  optical phonons~\cite{27,28} and acoustic phonons~\cite{31,32}, respectively, with $\tau_{\varepsilon} \propto (T_0/\hbar\omega_0)^2\exp(\hbar\omega_0/T_0)$ and
$\nu \propto T_0$. Here $\hbar\omega_0 \simeq 200$~meV is the optical phonon energy.
This implies that the characteristics obtained above might be substantially modified for GC-FET detectors operating at low temperatures. The latter is beyond the scope of this paper and requires a separate study.

\section{Conclusion}
We evaluated  the performance of the hot-electron GC-FET bolometers with graphene channel and the composite h-BN/b-P/h-BN gate BL and showed
that these bolometers can exhibit high values of the responsivity to the THz radiation modulated by signals of dozens GHz at room temperature.
The predicted high performance of the GC-bolometers might encourage the fabrication of these devices and their characterization.
%A conclusion section is not required. Although a conclusion may review the 
%main points of the paper, do not replicate the abstract as the conclusion. A 
%conclusion might elaborate on the importance of the work or suggest 
%applications and extensions. 

%\section*{Author declarations}

%The authors declare no conflict of interest.

\end{document}